\documentclass[reprint,floatfix,twocolumn,superscriptaddress]{revtex4-2}

\usepackage{amsmath}
\usepackage{amssymb}
\usepackage{hyperref}
\usepackage{float}
\usepackage{graphicx}

\newcommand{\Fref}[1]{Fig.~\ref{#1}}
\newcommand{\Tref}[1]{Tab.~\ref{#1}}



\begin{document}

\title{
A Millimeter-Wave Superconducting Qubit
}

\author{Alexander Anferov}
\email{aanferov@uchicago.edu}
\affiliation{James Franck Institute, University of Chicago, Chicago, Illinois 60637, USA}
\affiliation{Department of Physics, University of Chicago, Chicago, Illinois 60637, USA}

\author{Fanghui Wan}
\affiliation{Department of Applied Physics, Stanford University, Stanford, California 94305, USA}

\author{Shannon P. Harvey}
\affiliation{Department of Applied Physics, Stanford University, Stanford, California 94305, USA}
\affiliation{SLAC National Accelerator Laboratory, Menlo Park, CA, 94025 USA}

\author{Jonathan Simon}
\affiliation{Department of Applied Physics, Stanford University, Stanford, California 94305, USA}
\affiliation{Department of Physics, Stanford University, Stanford, California 94305, USA}

\author{David I. Schuster}
\email{dschus@stanford.edu}
\affiliation{Department of Applied Physics, Stanford University, Stanford, California 94305, USA}
\affiliation{SLAC National Accelerator Laboratory, Menlo Park, CA, 94025 USA}

\begin{abstract}
Manipulating the electromagnetic spectrum at the single-photon level is fundamental for quantum experiments.
In the visible and infrared range, this can be accomplished with atomic quantum emitters, and with superconducting qubits such control is extended to the microwave range (below 10~GHz).
Meanwhile, the region between these two energy ranges presents an unexplored opportunity for innovation.
We bridge this gap by scaling up a superconducting qubit to the millimeter-wave range (near 100~GHz).
Working in this energy range greatly reduces sensitivity to thermal noise compared to microwave devices, enabling operation at significantly higher temperatures, up to 1~K. 
This has many advantages by removing the dependence on rare $^3$He for refrigeration, simplifying cryogenic systems, and providing orders of magnitude higher cooling power, lending the flexibility needed for novel quantum sensing and hybrid experiments.
Using low-loss niobium trilayer junctions, we realize a qubit at 72~GHz cooled to 0.87~K using only $^4$He.
We perform Rabi oscillations to establish control over the qubit state, and measure relaxation and dephasing times of 15.8 and 17.4~ns respectively.
This demonstration of a millimeter-wave quantum emitter offers exciting prospects for enhanced sensitivity thresholds in high-frequency photon detection, provides new options for quantum transduction and for scaling up and speeding up quantum computing, enables integration of quantum systems where $^3$He refrigeration units are impractical, and importantly paves the way for quantum experiments exploring a novel energy range.
\end{abstract}

\maketitle
\section{Introduction}
Superconducting qubits are well-known as a promising quantum computing platform \cite{kjaergaard2020qubitReview}, but are also an invaluable tool for exploring electromagnetic phenomena with extreme sensitivity, as they can directly detect and manipulate electromagnetic signals at the quantum level \cite{akash,Besse2018spd,Inomata2016spd,Lescanne2020spd,Yin2013catch,Lotkhov2019spd100GHz,Houck2007photonsource}. 
In contrast to atomic systems, circuit parameters are easily adjusted, making them particularly useful for interacting with other types of quantum systems \cite{clerk2020hybrid,xiang2013hybridrev,Raimond2001haroche}, enabling new kinds of hybrid quantum experiments and sensors.
Thus far, superconducting qubits have operated between 100~MHz -- 10~GHz, leaving the millimeter-wave spectrum (near 100~GHz) unexplored.

Increasing qubit frequencies expands the energy range available to current quantum experiments, providing tools for studying new regions of the electromagnetic spectrum.
A high-frequency quantum system could help reach unprecedented sensitivity for millimeter-wave astronomical detection \cite{Monfardini2011mkid, Pirro2017bolometers,Ulbricht2021mkids,tucker1985millimeterrev}, and for studying unpaired electron interactions in molecular systems \cite{aslam2015esr,Kubo2012qubitesr,vasilyev2004mmesr,Drost2022scanesr}.
Millimeter-wave quantum circuits could also interact with new types of quantum emitters \cite{clerk2020hybrid,xiang2013hybridrev}, including neutral Rydberg atoms \cite{kumar2023transduct, chatterjee2023quantum,Raimond2001haroche}, and molecular qubits with higher frequency transitions \cite{Han2018molecular,Gaita-Arino2019molecular}.
Many of these experiments pose additional challenges by exposing qubits to magnetic fields or direct optical illumination \cite{kumar2023transduct, wang2022transduct, mirhosseini2020transduct, jiang2020transduct}, so the improved thermal resilience of higher-frequency qubits has additional practical advantages, reducing hybrid experiment complexity and improving performance at higher temperatures.
This could even allow superconducting quantum experiments in airborne or spaceborne payload environments \cite{Ulbricht2021mkids,Ade2020balloon,Masi2019balloon,Cataldo2019uspec} where cooling power is limited \cite{Chen2024spaceCooling}.

The benefits of higher temperature similarly translate to superconducting quantum computing, both in managing the heat load from increasing numbers of control signals \cite{Krinner2019qubitLines} in ever-growing quantum circuits \cite{deLeon2021qubitmaterials,Siddiqi2021qubitmaterials,kjaergaard2020qubitReview} and the heat from quantum processors integrated with digital logic \cite{liu2023sfqControl,leonard2019sfqControl,mcdermott2014sfqControl,Brock2000sfq100GHz}.
Millimeter-wave qubits could also reduce hardware overhead for superconducting quantum interconnects between cryostats \cite{pechal2017millimeter,magnard2020link,Multani2024links}, or reduce the energy difference required in optical transduction \cite{kumar2023transduct} for long-distance quantum communication.

\begin{figure*}
\centering
\includegraphics[width=6.67in]{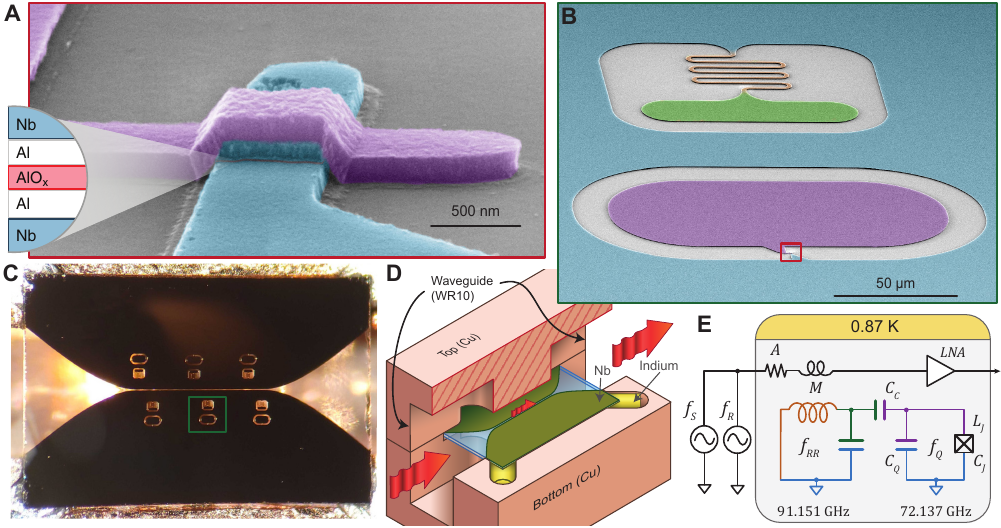}
\caption{
Device geometry.
(a) Composite scanning electron micrograph of the Nb/Al/AlO$_x$/Al/Nb Josephson junction at the heart of the qubit. A niobium wiring layer (purple) contacts the top of the junction, and the temporary SiO$_2$ scaffold is dissolved away.
(b) Scanning electron micrograph of a readout resonator (top, green) which is capacitively coupled to the qubit capacitor (bottom, purple).
(c-d) Photograph of several qubits and readout resonators coupled to a tapered finline transmission structure, along with a diagram illustrating how the structure couples signals to and from rectangular waveguides used for measurement.
(e) Simplified circuit representation of the experiment, where the readout resonator-qubit structure is inductively coupled to the readout and qubit control signals. The assembly is cooled to 0.87~K in a $^4$He adsorption refrigerator.
\label{fig1}}
\end{figure*}

A high-frequency superconducting qubit requires both low-loss circuit materials as well as a source of strong nonlinearity.
Recent developments have brought high coherence to linear systems in the millimeter-wave band, both as three-dimensional cavities \cite{kumar2023transduct,Kuhr2007cavity,Suleymanzade2020cavity} as well as in planar circuits \cite{anferov2024taper,Shan2024tlsloss,Hahnle2020cpwradiation,u-yenWollack2017}.
The nonlinearity presents a greater challenge, since the higher operating frequencies and temperatures break Cooper pairs \cite{Liu2024pairbreaking} in standard aluminum Josephson junctions used in microwave circuits \cite{kjaergaard2020qubitReview}, causing catastrophic dissipation from quasiparticles \cite{mattis1958bardeen,catelani2011quasiparticle,catelani2014quasiparticle}.
Alternate effects such as kinetic inductance \cite{Faramarzi2021kineticon, Winkel2020graalmon,Rieger2023graalfluxmon,Joshi2022ktin} in higher critical temperature ($T_c$) superconductors have been proposed for realizing a millimeter-wave qubit \cite{Faramarzi2021kineticon}: recently this has been used to demonstrate a weak nonlinearity \cite{anferov2020mmKI,stokowskiSafavi-Naeini2019,Joshi2022ktin}.

We instead address this challenge with a niobium Josephson junction \cite{morohashi1987nbjrev,gronberg2017swaps,anferov2023nbjj} recently used in qubits up to 24~GHz \cite{anferov2024kband}.
We investigate and demonstrate control of a transmon qubit at 72~GHz, and establish the viability of a millimeter-wave quantum system by operating at temperatures near 1~K, cooled without using rare $^3$He.

\section{Millimeter-wave Qubit Design}

The maximum operating frequency of a Josephson junction depends on the energy gap of its primary superconductor.
While niobium provides the necessary high energy gap \cite{novotny1975nbGap,turneaure1968nbGap} for millimeter-wave work, its oxides are lossy and imperfect insulators \cite{verjauw2021nbOxide,premkumar2021nbOxide} leading to very poor natural junction tunnel barriers. Therefore we utilize a thin layer of oxidized aluminum at the interface to form an aluminum oxide tunnel barrier, which has a clean interface with low leakage and loss. 
Through the superconducting proximity effect, this trilayer junction (illustrated in \Fref{fig1}a) is able to combine niobium’s superconducting gap and electrical properties with an aluminum oxide tunnel barrier, which has a clean interface with low leakage and loss.

These niobium trilayer junctions are fabricated on a 100$~\mu$m-thick sapphire wafer using a process detailed in Ref. \cite{anferov2023nbjj}.
Following the deposition of the trilayer stack, an etch defines the bottom layer geometry, and a self-aligned sacrificial spacer \cite{gronberg2017swaps} scaffold made of SiO$_2$ allows a final layer of niobium (purple in \Fref{fig1}a) to contact the top of the trilayer only.
This final niobium layer is etched along with the top half of the trilayer stack to define a junction in the overlapping region.

The operating range of a Josephson junction is set by its critical current density $J_c$ and its intrinsic capacitance $C_J$, which together define the self-resonant plasma frequency $\omega_p=1/\sqrt{L_J C_J}=\sqrt{2 e J_c \epsilon/(\hbar d)}$, where the tunnel barrier has dielectric $\epsilon$ of thickness $d$  \cite{kanter1970current}.
Heating niobium Josephson junctions to over $\sim100~^\circ$C is known to decrease $J_c$ \cite{Migacz2003nbAnneal,anferov2023nbjj,morohashi1987nbjrev}, so to preserve its properties we avoid exposing the junction to elevated temperatures at anytime throughout the fabrication process.
For this reason, the SiO$_2$ spacer material is grown at the comparatively low temperature of $90~^\circ$C in order to keep the plasma frequency as high as possible.
For our junctions, we find that $J_c=1.4~\text{kA}/\text{cm}^{2}$: combined with the specific electrode capacitance and a barrier thickness just over 1~nm \cite{liu2023unveiling,morohashi1987nbjrev} we estimate that this corresponds to a junction plasma frequency of about 99~GHz.
We note that higher processing temperatures (up to $230~^\circ$C) which have subsequently been found to produce better-quality scaffold material could be used instead without significant reduction in $J_c$ \cite{anferov2024kband, morohashi1987nbjrev}.
A final ammonium-fluoride etch dissolves the lossy spacer material along with some niobium surface oxides.
However as later revealed by cross-sectional images (see \Fref{fig:S10}), this etch  does not completely remove the spacer material, and also dissolves part of the aluminum, exposing more niobium surface area for re-oxidation.

\begin{figure*}
\centering
\includegraphics[width=6.67in]{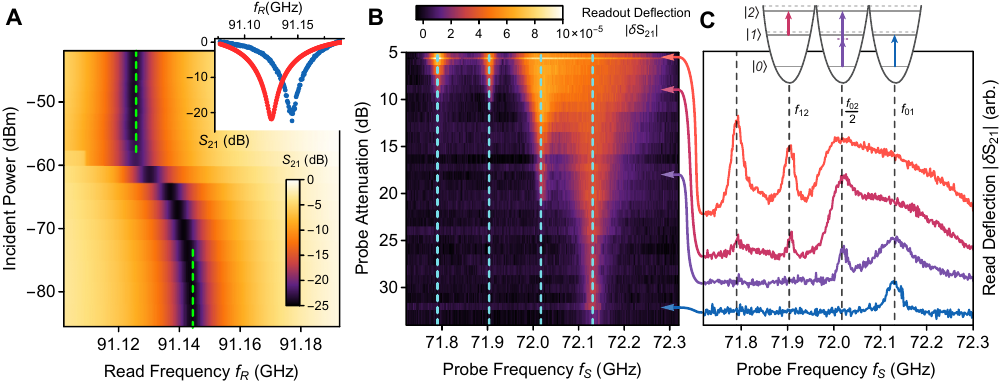}
\caption{
Qubit spectroscopy.
(a) Single tone resonator spectroscopy as a function of applied power. While the resonant frequency at low powers is dressed by the presence of the qubit, above a critical power, the qubit decouples from the resonator, which returns to its bare frequency.
Line cuts of the low and high-power transmission magnitude are shown in the inset.
(b) Readout resonator transmission as a function of applied probe tone frequency, shown for increasing probe power.
Line cuts for indicative qubit drive powers are shown in (c).
At low powers (blue) a single peak is observed when the pulse is resonant with the qubit frequency ($f_{01}=72.1~$GHz).
As power increases, the linewidth of this transition increases, and additional peaks appear from excitations into higher qubit levels through many-photon excitations ($f_{02}/2$ etc).
These features have a spacing of $\alpha/2 = (f_{01}-f_{12})/2=114~$MHz.
Based on the amplitude ratio of the $f_{01}$ and $f_{12}$ features at low power, we estimate an upper bound of 6.33\% for the $|1\rangle$ state population, indicating a qubit temperature below $1.287~$K.
\label{fig2}}
\end{figure*}

The qubit and readout resonator geometry (shown in \Fref{fig1}a) is defined using optical lithography during the junction fabrication process.
The junction is larger than those used for conventional microwave qubits \cite{kjaergaard2020qubitReview} with a nominal area of $0.56~\mu\text{m}^2$, providing a significant amount of capacitance (approximately 45~fF).
This junction is shunted with a 39~fF capacitor (purple electrode in \Fref{fig1}) to form a transmon qubit \cite{koch2007cpb}: this structure is 150 microns wide, resulting in a small device footprint.
The qubit capacitor is in turn coupled to the capacitor (green electrode) of a millimeter-wave lumped-element resonator \cite{anferov2024taper} for dispersive readout.

The qubit and readout resonator interact with measurement and control signals through a tapered coupling structure (described in Ref. \cite{anferov2024taper}) shown in \Fref{fig1}c-d, which transforms the high-frequency signals from a copper rectangular waveguide to and from an on-chip slotline waveguide.
Analagously to coplanar waveguide feedlines used in microwave devices \cite{barends2013xmon, kjaergaard2020qubitReview}, confining the electromagnetic fields of the signal to a slotline allows us to decouple the qubit from the feedline while maintaining resonator coupling.
The contrast between resonator-feedline and qubit-feedline coupling is further enhanced by inductively coupling the readout resonator to the slotline, as shown in \Fref{fig1}e, which enables stronger readout coupling strengths compared to capacitive coupling \cite{anferov2024taper} at a given distance from the slotline.
The tapered coupling structure also provides a ground plane for the qubit and readout resonator, helping reduce dissipation from dipole radiation \cite{brecht2015seam}.
The system can be modeled with the circuit diagram shown in \Fref{fig1}e.

\section{Qubit Energy Spectrum}
Establishing control over the qubit state requires first determining the system parameters, achieved using transmission measurements through the assembly in \Fref{fig1}d which is cooled to 0.87~K in a $^4$He adsorption refrigerator.
First we investigate the system with a single signal at frequency $f_R$.
At low power, we observe a dip in transmission near the dressed \cite{koch2007cpb,reed2010snap,bishop2010snap} resonant frequency of the readout resonator $f_\text{RR}$, as shown in the inset of \Fref{fig2}a.
As the applied signal power increases, the resonant feature shifts down in frequency due to inherited nonlinearity from the coupled qubit.
Despite higher kinetic inductance at millimeter-wave frequencies, this nonlinearity is orders of magnitude stronger than kinetic inductance in identical niobium \cite{anferov2024taper} or even niobium nitride resonators \cite{anferov2020mmKI} confirming the presence of a strongly nonlinear element.
At a sufficiently high power the resonance arrives at the bare readout resonator frequency as the qubit saturates \cite{reed2010snap,bishop2010snap}.
When combined with the qubit-resonator frequency detuning $\Delta_{qr}$,this difference between the dressed and bare readout resonator frequencies $g^2/\Delta_{qr}$ determines the bare coupling strength $g/2\pi = 607.9~$MHz in the system Hamiltonian, which is reasonably consistent with simulations \cite{minev2021epr}.

While monitoring the readout resonator ($f_R$) at low-power, simultaneously adding a probe signal ($f_S$) reveals the energy spectrum of the qubit.
In \Fref{fig2}b-c we show change in readout signal $\delta S_{21} = |S_{21}(P_s)-S_{21}(0)|$ as a function of probe frequency and power $P_s$.
At low powers, we observe a deflection in transmission at the bare qubit frequency $f_{01}=72.137~$GHz as the qubit is excited.
The linewidth of this feature increases with applied power: this scaling can be used to estimate the qubit coherence properties \cite{schuster2005ac} (see \Fref{fig:S7}), predicting qubit dephasing times on the order of 20~ns. 
Stronger probe powers reveal the higher energy states of the qubit through two-photon processes $f_{02}=(f_{01}+f_{12})/2$ and excited state transitions $f_{12}$.
The spacing of these features (228~MHz) is approximately equal to the charging energy of the qubit island $E_c$, which sets the qubit anharmonicity $\alpha \equiv f_{12}-f_{01}$.
Notably, while other device parameters are scaled up significantly compared to typical values (see \Tref{tab:qubitparams}), the chosen anharmonicity is similar to conventional microwave qubits \cite{paik2011qubit3d,koch2007cpb,barends2013xmon}, resulting in comparable limits to the speed of qubit control operations.
In future designs, higher anharmonicity could be achieved with a smaller junction area or by reducing qubit capacitance.

\bgroup
\def\arraystretch{1.25}
\begin{table}
\caption{\label{tab:qubitparams}Summary of device parameters.}
\noindent
\begin{center}
\begin{tabular}{c c c c c c c c}
\hline
$E_J/2\pi$&$J_c$&$A_J$&$C_Q$ (sim)\\
\hline
2.871~THz&1.43~$\text{kA}/\text{cm}^2$&$0.56~\mu\text{m}^2$&39~fF\\
\hline
\noalign{\vskip 1mm}
\hline
$\omega_{RR}/2\pi$&$\gamma_{RR}/2\pi$&$g^2/2\pi\Delta$&$\omega_Q/2\pi$\\
\hline
91.151~GHz&84.281~MHz&-19.44~MHz&72.137~GHz\\
\hline
\noalign{\vskip 1mm}
\hline
$\Delta$&$E_C/2\pi$&$g/2\pi$& $\chi/2\pi$\\
\hline
-19.0143~GHz&228~MHz&607.979~MHz&-0.230~MHz\\
\hline
\end{tabular}
\end{center}
\end{table}
\egroup

\section{Time-Domain Measurements}
\begin{figure}
\centering
\includegraphics[width=3.1in]{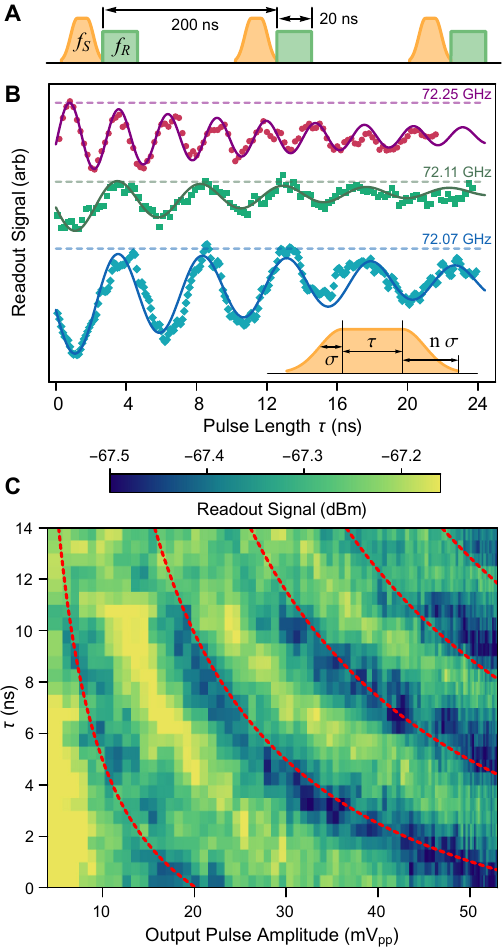}
\caption{
Rabi oscillations.
(a) The device is repeatedly probed with shaped pulses near the qubit frequency followed by a pulsed phase-coherent signal at the readout frequency, while continuously acquiring the filtered output signal.
(b) The qubit pulse induces oscillations between its ground and excited states, visible by measuring the readout resonator state while varying pulse length $\tau$ (with $\sigma=1.5~$ns as shown in the inset).
For each measurement the ground state is indicated with a dashed line.
Near the qubit frequency, the Rabi oscillation contrast increases while the frequency decreases.
(c) Rabi frequency should increase with pulse amplitude, demonstrated by measuring oscillations as function of calibrated pulse amplitude and length $\tau$, with darker colors corresponding to larger readout resonator shifts.
Dashed red lines mark contours of integer $\pi$ pulses where $\sigma\Omega=n\pi$.
At very high pulse powers the Rabi frequency deviates from the expected linear relationship.
\label{fig3}}
\end{figure}
With the system parameters established, we investigate the qubit behavior in the time domain with the goal of calibrating a control pulse.
Implementing this measurement at millimeter-wave frequencies presents new hardware challenges, further complicated by the need for rapid experiments with individual durations below 40~ns due to estimated coherence times (see \Fref{fig:S7}), while also maintaining relative signal phase between experiments.
We solve this by digitally switching the output of the phase-locked measurement signals from the network analyzer used for previous continuous measurements (see \Fref{fig:S1}).
This output is combined with a qubit control signal, generated by mixing a synthesized 6~GHz microwave waveform to millimeter-wave frequencies in a heterodyne fashion, and synchronized with the measurement pulses as shown in \Fref{fig3}a.
Individual experiments are repeated every 200~ns during data acquisition, which coherently averages pulsed transmission through the readout resonator.

Driving the qubit with a control pulse results in Rabi oscillations between its ground and excited states depending on the pulse area, reaching a minimum frequency and maximum amplitude as the pulse is brought on resonance with the qubit transition frequency $f_{01}$. 
We use a constant amplitude control pulse with $\sigma=1.5~$ns Gaussian edges, as a compromise between constant oscillation rates while minimizing pulse bandwidth.
We summarize the oscillations for different pulse frequencies in \Fref{fig3}b, with a maximum Rabi frequency $\Omega_0 = 208~$MHz (see \Fref{fig:S8} for additional frequency dependence).
These oscillations decay at longer pulse lengths due to qubit dephasing.

To verify the time-dependence of the Rabi oscillations, we also repeat this measurement at the optimal pulse frequency while varying pulse length $\tau$ and plot the results in \Fref{fig3}c.
We verify that the Rabi oscillation frequency $\Omega$ increases at larger pulse amplitudes, corresponding to more closely spaced fringes.
When accounting for the Gaussian sections of the pulse, this scaling is approximately proportional to $\tau$, with some deviation at large amplitudes, where the microwave waveform amplitude exceeds the linear regime of the millimeter-wave mixer and the effective Rabi frequency is comparable to the qubit anharmonicity.
The fringe contours from this measurement help calibrate integer rotations between the ground and excited states of the qubit.
From these we select a control $\pi$ pulse with $\tau=2~$ns and an amplitude corresponding to $\Omega/2\pi=68~$MHz (well below the anharmonicity) and a $\pi/2$ pulse with the same duration but half the amplitude.

\section{Coherence Properties}
\begin{figure}
\centering
\includegraphics[width=3in]{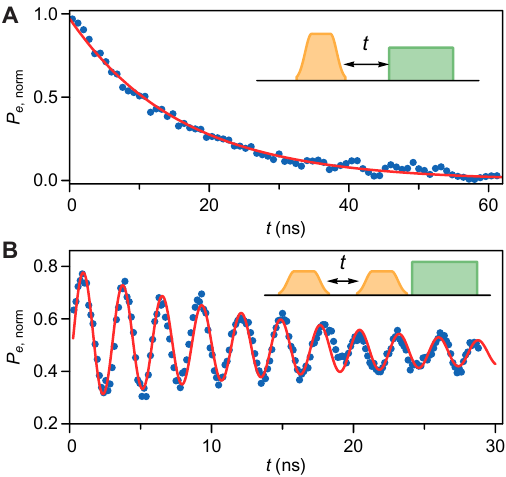}
\caption{
Qubit coherence properties.
(a) Relaxation time measured by fitting the normalized excited state population (readout deflection relative to maximum Rabi fringe contrast) as a function of measurement delay.
The red line is a exponential fit with a characteristic time $T_1=15.849~$ns.
(b) Ramsey dephasing experiment, consisting of two half-amplitude control pulses separated by a delay.
The phase of the second pulse is advanced at 320~MHz relative to the first, resulting in decaying oscillations with a characteristic dephasing time $T_2^*=17.466~$ns.
Since the control pulses lengths are comparable to coherence times, we observe reduced fringe contrast and phase offset.
Drift and shifted oscillation frequency suggest imperfect control pulse calibrations.
\label{fig4}}
\end{figure}
Having established control over the qubit state, we can now study qubit relaxation time and dephasing time in the time domain, which dictate qubit limitations and act as sensitive probes for millimeter-wave loss channels.
We measure relaxation time by placing each qubit in its excited state with the calibrated $\pi$ pulse length above and comparing the measured qubit population after time $t$ to the initial maximum value, as shown in \Fref{fig4}a.
Fitting this exponential state decay yields the characteristic relaxation time $T_1$, which we find to be 15.849~ns, which corresponds to a unit-less qubit quality factor $Q_1 = \omega_Q T_1 = 7.18\times10^3$.

Next we perform Ramsey interferometry as summarized by the interference fringe shown in \Fref{fig4}b, which decays with the characteristic dephasing time $T_2^*$. 
From this experiment we extract a dephasing time of 17.466~ns.
Due to the length of a single pulse relative to the qubit coherence time, the oscillations begin with reduced contrast, since the qubit has already begun to decohere.
From the relationship between $T_2^*$ and the coherence time $1/T_2^{*} = 1/T_\phi + 1/2 T_1$, we calculate a pure dephasing rate $T_\phi=38.90~$ns, suggesting the presence of significant noise reaching the device.
This suggests that the measured coherences have not yet approached fundamental limits, and identifies input filtering and thermalization as a direct path for improvement.

\section{Discussion}
Our device energies are substantially higher than previously measured superconducting devices, providing insight into the nature of single-photon coherence in a new frequency regime.
The quality factors measured for our device are commensurate with early microwave qubit demonstrations \cite{houck2008controlling}, and we anticipate significant future improvements as loss channels are addressed.
Since the qubit junction contains a significant proportion of the qubit capacitance, we estimate that the coherence is primarily limited by material losses in the trilayer junction (see \Fref{fig:S10}).
Promisingly, an improved spacer deposition method \cite{anferov2024kband} has subsequently demonstrated improved junction losses at elevated microwave frequencies while preserving high junction critical current densities needed for millimeter-wave devices.
While simulations suggest our qubit is not likely limited by spontaneous emission \cite{purcell1995spontaneous,Houck2008purcell,Sete2014purcellnonlin} through the readout resonator or taper structure  (see \Fref{fig:S9}), radiation loss through the in-plane seam of the sample housing \cite{brecht2015seam, anferov2024taper} could be another significant decoherence channel.
This outlines several clear pathways for improving coherence in future millimeter-wave qubits, suggesting that any fundamental millimeter-wave coherence limits are not yet reached.

We also note that our current measurement hardware and qubit design is far from optimized, necessitating extensive averaging of experiments to reconstruct the qubit state.
Furthermore, our pulse generation electronics do not successfully filter the strong byproduct signals generated in the pulse up-conversion process (see \Fref{fig:S3}), as evidenced by ac-Stark shifts \cite{Schneider2018multiStark} in the pulsed spectroscopy qubit transitions relative to those measured in \Fref{fig2}.
This added power and noise likely significantly limits our qubit performance \cite{dumas2024ionization,Shillito2022ionization,schuster2005ac,Gambetta2006broadening}, but could be remedied with improved filtering techniques.
Our measurements also have limited qubit state visibility, because readout measurements currently take longer than a coherence time.
Switching to a high-speed digitized readout scheme \cite{stefanazzi2022qick} could greatly improve measurement efficiency: combined with increased anharmonicity or optimized pulse shapes \cite{Chow2010drag,Werninghaus2021qoc} and faster gates with improved fidelities, this would unlock the full high-speed capabilities of millimeter-wave qubits.

\section{Conclusion}
We have demonstrated control over a superconducting artificial quantum system operating at higher frequencies and temperatures than ever before.
Our 72~GHz qubit has sufficiently high transition energies to function when cooled to 0.87~K with simpler $^4$He refrigeration methods.
This device highlights the possibilities of superconducting quantum experiments in new environments, whether as sensitive detectors or efficient repeaters in a quantum network, and provides new options for addressing thermal challenges in quantum processors or reducing experimental complexity in hybrid quantum experiments.
Moreover, extending circuit performance to higher frequencies expands the energy scale of quantum systems that can be modelled with superconducting qubits.
Niobium-based junctions can operate at even higher frequencies in the sub-THz range; already our device demonstrates the possibility of higher-frequency, higher-temperature qubits, opening new pathways and new applications for superconducting quantum devices in a novel energy regime.

\section*{Acknowledgments}
The authors thank P. Duda for assistance with fabrication development and A. Oriani for cryogenic design asistance.
This work is supported by the U.S. Department of Energy Office of Science National Quantum Information Science Research Centers as part of the Q-NEXT center, and partially supported by the University of Chicago Materials Research Science and Engineering Center, which is funded by
the National Science Foundation under Grant No. DMR-1420709.
This work made use of the Pritzker Nanofabrication Facility of the Institute for Molecular Engineering at the University of Chicago, which receives support from Soft and Hybrid Nanotechnology Experimental (SHyNE) Resource (NSF ECCS-2025633). Part of this work was performed at the Stanford Nano Shared Facilities (SNSF), supported by the National Science Foundation under award ECCS-2026822.

\nocite{dataset}

\appendix


\section{Device Fabrication}
\label{appendix:fab}
\begin{figure*}[htb]
\centering
\includegraphics[width=6.67in]{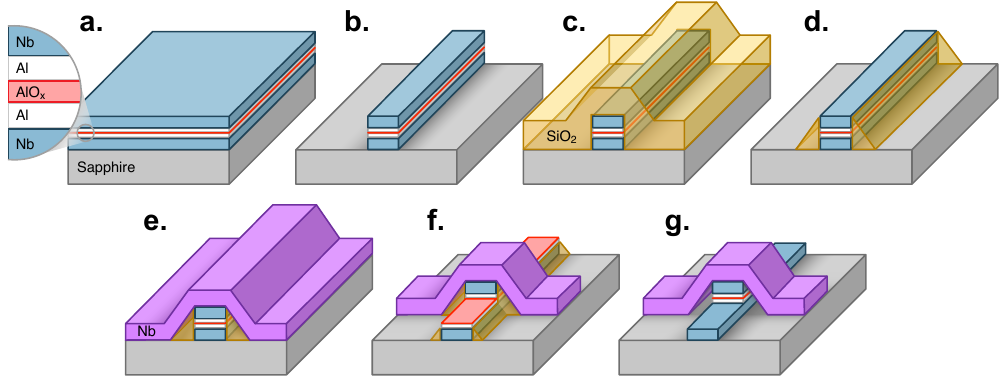}
\caption{
Junction fabrication process.
(a) Trilayer is deposited and oxidized in-situ. (b) First layer is etched with a chlorine RIE. (c) SiO$_2$ is grown isotropically at $90^\circ~$C using HDPCVD. (d) Sacrificial spacer is formed by anisotropic etching with fluorine chemistry. (e) Surface oxides are cleaned in vacuum and niobium wiring layer (purple) is deposited. (f) Second junction finger (and other circuit elements) are defined by a fluorine plasma etch selective against Al. (g) Final devices undergo a wet etch to further remove SiO$_2$ and exposed Al.
\label{fig:fab}}
\end{figure*}
\begin{table*}
\noindent
\begin{center}
\begin{tabular}{|c||c|c|c|c|c|c|c|c|c|c|c|c|}
\hline
&T($^\circ$C)&Pressure &
ICP/Bias& Cl$_2$ & BCl$_3$ &Ar &CF$_4$ &CHF$_3$ &SF$_6$ &O$_2$ &etch time &etch rate\\
\hline
Etch 1 &$20\pm0.1$ &5 mT &400~W / 50~W &30 &30 &10 &- &- &- &- &50-60s&$\sim4.5~\text{nm}/s$\\
Etch 2 &$20\pm0.1$ &30 mT &500~W / 60~W &- &- &10 & 30& 20&- &- &120-140~s&$\sim2~\text{nm}/s$\\
Etch 3 &$20\pm0.1$ &5 mT &400~W / 60~W  &- &- &7 &- &40 &20 &4 &65-90~s&$\sim4.5~\text{nm}/s$\\
\hline
\end{tabular}
\end{center}
\caption{Plasma etch parameters used in the ICP-RIE etches described in the process. Etches are performed in an Apex SLR ICP etcher. Gas flows are listed in sccm.}
\label{tab:etches}
\end{table*}
The fabrication process is adapted from Ref. \cite{anferov2023nbjj} for use with thin sapphire wafers.
100~$\mu$m-thick C-plane polished sapphire wafers undergo an ultrasonic clean in organic solvents (toluene, acetone, methanol, isopropanol), then annealed at in a nitrogen atmosphere (We note that nitrogen annealing has since been shown to damage the sapphire.) at 1100~$^\circ$C for two hours and allowed to cool to room temperature. 
The wafers are treated to a second ultrasonic solvent clean, then etched in a piranha solution heated to 40~$^\circ$C for 2 minutes.
Immediately following, the wafers are loaded into a Plassys MEB550S electron-beam evaporation system, where they undergo a dehydration bake at $>$200~$^\circ$C under vacuum for an hour.
When a sufficiently low pressure is reached ($<{5\times10^{-8}}$~mBar), titanium is electron-beam evaporated to bring the load lock pressure down even further.
While rotating the substrate, the trilayer is now deposited by first evaporating 80~nm Nb and 8~nm Al (at a 10 degree angle).
The aluminum is lightly etched with a 15~mA 400~V Ar$^+$ beam for 10~s, then oxidized with a mixture of $15\%$ O$_2$:Ar at 3~mBar for 3 minutes.
After lowering the vacuum pressure with titanium, we deposit 3~nm of Al and 150~nm of Nb to finish the trilayer.
After cooling for a few minutes, the top surface is oxidized at 3~mBar for 30~s.

The wafers are mounted on a silicon handle wafer using AZ703 photoresist cured at 115~$^\circ$C, then coated with 1~$\mu$m of AZ MiR 703 photoresist and exposed with a 375~nm laser in a Heidelberg MLA150 direct-write system.
The trilayer is now etched in a chlorine inductively coupled plasma reactive ion etcher (Etch 1 in Table \ref{tab:etches}), then quenched with DI water to dilute adsorbed HCl.
The remaining photoresist is thoroughly dissolved in a mixture of 80~$^\circ$C n-methyl-2-pyrrolidone (NMP), which also releases the substrate from the handle wafer.
The wafer is ultrasonically cleaned with acetone and isopropanol, then mounted on a silicon handle wafer with Crystalbond 509 wax.
The SiO$_2$ spacer is now grown with a SiH$_4$ O$_2$ and Ar plasma, with the substrate heated to 90~$^\circ$C, then etched with a fluorine reactive ion etch (Etch 2 in Table \ref{tab:etches}) to form the spacer structure.
Wafers are now removed from the handle wafer by melting the wax, ultrasonically cleaned, then immediately placed back in the deposition chamber, where they are gently heated to 50~$^\circ$C for 30 min to remove remaining volatiles.

The surface oxides are etched with a 15~mA 400~V Ar$^+$ beam for 5~min, and titanium is used to lower the vacuum pressure before depositing 160~nm of Nb to form the wiring layer.
The surface is passivated with $15\%$ O$_2$:Ar at 3~mBar for 30~s.
A second lithography step defines the rest of the circuit on the wiring layer, which is formed through a fluorine reactive ion etch (Etch 3 in Table \ref{tab:etches}).
The etch time is calculated for each wafer based on visual confirmation when the bare wiring layer is etched through.
With the junctions, resonant structures and taper geometry formed, the remaining resist is now fully dissolved in 80~$^\circ$C NMP, ultrasonically cleaned, then the wafer is diced into $2.2\times$3.2~mm chips with a layer of protective photoresist.
This coating is now dissolved in 80~$^\circ$C NMP, and the chips are given a final ultrasonic clean with with acetone and isopropanol.
The residual silicon spacer is now dissolved by a short 10$~$s etch in a mixture of ammonium fluoride and acetic acid (AlPAD Etch 639), quenched in de-ionized water, then carefully dried from isopropanol to preserve the now partially suspended wiring layer.
The finished chips are packaged and cooled down within a couple hours from this final etch to minimize any NbO$_x$ regrowth from air exposure.

Following approximately one year of measurements (See Appendix \ref{appendix:c}) the chip was unmounted, re-etched with an additional $15$~s of ammonium fluoride and acetic acid, then immediately remounted with fresh indium and cooled back down for the measurements in the main text.

\section{Measurement Setup}
\subsection{Experiment Refrigeration}
\begin{figure*}
\centering
\includegraphics[width=5.0in]{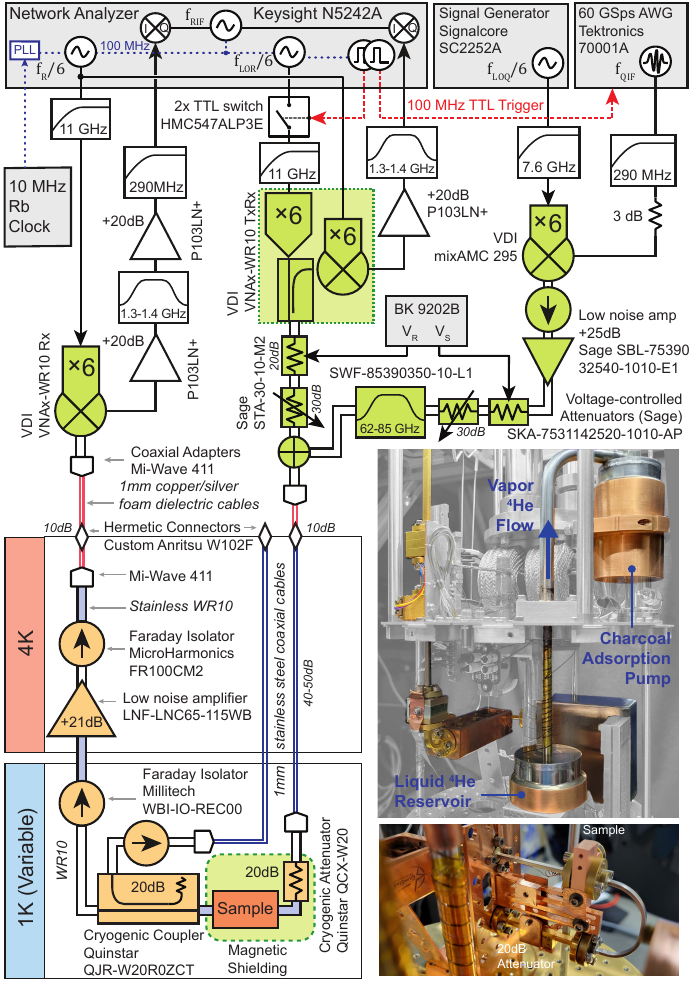}
\caption{
Schematic of the cryogenic millimeter-wave experiment setup. Green and orange shaded components represent room temperature and cryogenic millimeter-wave hardware respectively. Colored tabs show temperature stages inside the Helium-4 adsorption refrigerator, which reaches a base temperature of $0.86~$K.
Color photographs highlight the relevant hardware inside the fridge, as well as the refrigerator cooling mechanism.
The inset highlights components located inside the magnetic shield.
\label{fig:S1}}
\end{figure*}
\begin{figure*}[htb]
\centering
\includegraphics[width=6.5in]{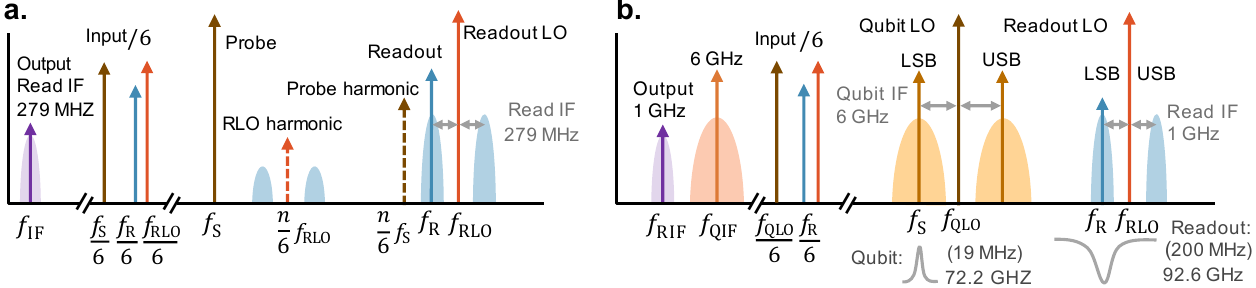}
\caption{
Frequency spectrum (not to scale) of the signals used. 
Continuous experiments are shown in (a) and pulsed experiments (b). The microwave signals are multiplied to produce desired millimeter-wave harmonics and other unwanted harmonics (dashed lines). These are omitted in (b) for brevity. Heterodyne conversion generates and is sensitive to both upper and lower sidebands (USB and LSB) on either sides of the local oscillator (LO). While the qubit image frequency (USB) is far enough away to be filtered, it is simpler to generate the readout signal directly with a multiplier, ensuring the device is measured with a single readout frequency. When mixed with the readout LO this produces a signal at the difference frequency (IF).
\label{fig:S2}}
\end{figure*}
All millimeter-wave characterization was performed in a custom built $^4$He adsorption refrigerator, which provides cooling through the evaporation of liquid $^4$He from a reservoir into an activated charcoal adsorption pump, as shown in the inset of \Fref{fig:S1}.
The system reaches a base temperature of $0.86~$K with a hold time of 3 hours.
The saturated adsorption pump can then be heated to refill the reservoir: this process takes 1.5 hours and resets the cooling mechanism, allowing for cycled cryogenic operation below 1~K.
The temperature of the sample is monitored and the experiments are synchronized to ensure data is only collected once the system has stabilized to within 10~mK of the base temperature.

\subsection{Millimeter-wave Measurements}
While in conventional microwave experiments pulses can now be generated and digitized directly, high-frequency measurements present additional challenges since signals need to be converted to and from the millimeter-wave band.
We approach this with a simple measurement method that avoids the need for filtering unwanted images by combining direct upconversion with heterodyne downconversion.
The hardware is outlined in \Fref{fig:S1} with relevant frequencies shown in \Fref{fig:S2}.
The key to achieving phase-coherent averaged measurements with this method lies in using microwave signals from a network analyzer generated from the same phase interal locked loop.
The microwave measurement signal ($12$-$19~$GHz) is sent into a frequency multiplier to generate signals in the millimeter-wave range ($75$-$115~$GHz).
For pulsed experiments, the readout signal is pulsed with a microwave switch (which has a $\sim5~$ns turn-on time).
Reflections of the upconverted signal are sampled to establish a phase reference measurement.
For continuous two-tone measurements the measurement signal is combined with a millimeter-wave probe signal generated with a multiplier.
Along with the desired signal $f_S$, this also produces harmonics evenly spaced by the microwave generating frequency ($f_S/6$) as shown in \Fref{fig:S2}a.
For pulsed qubit measurements the output is instead combined with the products of a heterodyne mixer, which maps a high-resolution microwave pulse to millimeter-wave frequencies.
The unwanted mixer images are amplified and filtered (we note that the filter used has only mild attenuation at the image frequency).
Note that for continuous measurements, the mixer-multiplier, amplifier and filters are replaced with a single multiplier.
Both qubit and readout signal powers are independently controlled by a combination of manual vane attenuators and voltage-controlled pin diode attenuators.
We note that the voltage-controlled attenuators have a relatively high noise temperature compared to their passive counterparts.

Thus far the combined signals to be sent into the device are waveguide TE$_{10}$ modes: we convert these to a $1~$mm diameter stainless steel and beryllium copper coaxial cable, which carries the signal to the $1~$K stage of the fridge, thermalizing mechanically at each intermediate stage, then convert back to a WR-10 waveguide which leads to the device under test. 
The cables and waveguide-cable converters have a combined frequency-dependent loss ranging from $38.6~$dB to $49.8~$dB in the W-Band, which is dominated by cable loss. 
The signal is further thermalized to $1~$K by a cryogenic 20 dB attenuator, which is mechanically anchored to the $1~$K stage of the refrigerator with copper mounts.
The copper sample mount, equivalent to a section of WR-10 waveguide, is also thermalized in a similar fashion, as shown in the inset of \Fref{fig:S1}.
These components are enclosed in a single-layer magnetic shield with high permeability (mu metal).

Having interacted with the sample through the taper coupling structure (detailed in Ref. \cite{anferov2024taper}), the output signal travels through the transmitted port of a cryogenic directional coupler (intended for reflection measurements but not used for this experiment, so the coupling port is blocked with an isolator) to a low noise amplifier.
Cryogenic faraday isolators minimize retro-reflections and prevent thermal radiation from leaking in on the output side, while still allowing good transmission.
After passing back outside the cryostat through custom-built hermetic adapters, the signal is downconverted and amplified for measurement.
The entire setup is summarized by \Fref{fig:S1} and \Fref{fig:S2}.

\section{Probe Signal Calibrations}
\begin{figure*}
\centering
\includegraphics[width=6.67in]{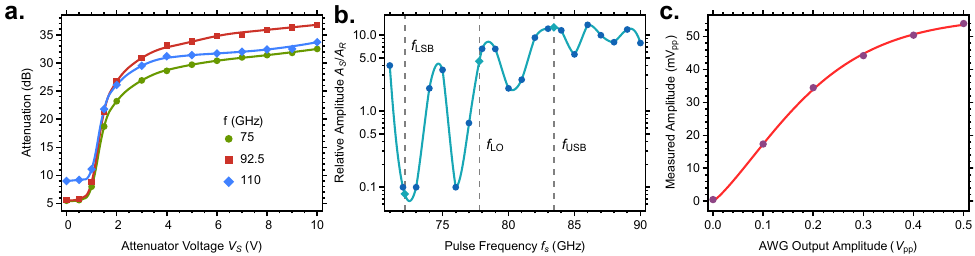}
\caption{
Probe signal calibrations.
(a) Qubit channel attenuation as a function of attenuator voltage, along with interpolations. The continuous output power is tunable across 20--30dB, but frequency-dependent and highly nonlinear with respect to applied voltage.
(b) Qubit pulse amplitude measured with an oscilloscope after downconverting with the readout mixer, normalized to the readout pulse amplitude at a fixed power.
Solid lines are interpolations used for approximating the qubit mixer-multiplier conversion loss, which shows significant ripples. We find an unfavorable 36~dB power discrepancy between the desired lower sideband and unwanted upper sideband (marked with dashed lines).
(c) Downconverted pulse amplitude measured near the qubit frequency as a function of microwave pulse amplitude, along with interpolation used in the main text, highlighting significant gain compression in the qubit pulse chain (primarily from the amplifier and mixer-multiplier).
\label{fig:S3}}
\end{figure*}
For readout and continuous qubit measurements, the millimeter-wave signals are generated using multipliers, which despite minor instabilities at lower powers \cite{kumar2023transduct} are relatively stable when operated in the saturation limit.
For this reason, the qubit microwave input power is held constant in the saturation limit, and the probe power $P_S$ is instead adjusted using a combination of manual attenuators for coarse adjustments and voltage-controlled pin diode attenuators for programmable sweeps.
In \Fref{fig:S3}a we plot a calibration of the attenuator insertion loss as a function of applied voltage, shown for several frequencies in the W-band.
The device has some frequency-dependent loss (most noticeable above 100~GHz), and consistent with pin diode attenuator design is highly nonlinear with respect to applied voltage.
Combined with a frequency-dependent calibration of the fridge line insertion loss and multiplier output power (see \cite{anferov2024taper,anferov2020mmKI}), interpolations of the attenuator calibration shown in \Fref{fig:S3}a can translate the control voltage $V_S$ and the probe frequency $f_S$ to the incident power arriving at the qubit.

For pulsed measurements the multiplier is replaced with a mixer-multiplier for heterodyne signal generation: the constant microwave signal previously used to generate the probe tone is now multiplied and used as the qubit local oscillator (LO) frequency (see \Fref{fig:S2}b) for mixing with a microwave pulse.
Unlike the multiplier, the output power of the qubit pulse (in this case the lower side band) now depends on the frequency-dependent conversion loss of the mixer-multiplier, the microwave image pulse frequency and amplitude, on top of the power and frequency of the LO signal.
Since calibrating the output for each of these variables is exhaustive, we instead calibrate the pulse amplitude seen by the qubit in-situ by measuring its amplitude after traversing the measurement setup with no amplification, filtering or attenuation (this also includes information about the insertion loss from the taper structure).
For this measurement, the readout LO is tuned to measure the qubit signal ($f_{RLO} = f_S + f_{RIF}$).
The resulting amplitude is summarized relative to the readout signal amplitude in \Fref{fig:S3}b.
Significant frequency-dependent amplitude variation can be observed in the qubit pulse amplitude, likely dominated by a combination of fluctuations in mixer conversion loss, insufficient mixer LO power, and impedance mismatches between physical components in the signal chain reflecting power back into the mixer.
Here we have attempted to address the latter by introducing an isolator and input attenuator to help reduce signal reflections (see \Fref{fig:S1}).
Unfortunately, our qubit frequency lines up with a minimum in multiplier efficiency, resulting in a qubit pulse signal approximately 100 times weaker than its upper sideband image (or approximately 36 dB lower in power).

Since the control pulse is actually the weakest of three signals generated by the mixer-multiplier, the output is amplified and filtered to produce sufficient power for controlling the qubit.
To minimize added noise injected into the qubit we use a low-noise amplifier.
We note that the filter used has a 3~dB cutoff at 85~GHz, so is only sufficient to slightly filter out the unwanted image frequency.
The low gain compression powers of the mixer-multiplier and the output amplifier respectively result in noticeable amplitude distortion, as shown in \Fref{fig:S3}c.
The qubit control pulse amplitude is recorded as a function of the AWG amplitude (determined by a combination of AWG output gain control between 0.25--0.5V and digital amplitude reduction) near the qubit frequency (to minimize frequency-dependent effects).
An interpolation allows us to to translate the generated pulse gain to its actual amplitude, which is used for rescaling amplitude data in the main text and Appendix \ref{appendixF}.

\section{Broadband Characterization}
\subsection{Single-tone Measurements}
\label{appendix:c}
\begin{figure*}[htb]
\centering
\includegraphics[width=5.0in]{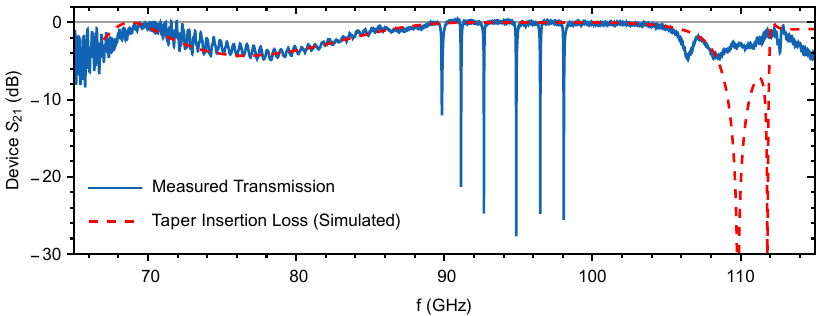}
\caption{
Measured transmission through the qubit sample. Here the wiring and millimeter-wave conversion circuitry is de-embedded.
Sharp dips in transmission occur at the readout resonance frequencies.
Broader features in the transmission spectrum are well-modeled by the simulated insertion loss from the tapered coupling structure.
\label{fig:S4}}
\end{figure*}
Measuring the qubit-resonator system begins with broadband transmission parameter measurements, which determine readout resonator frequencies and insertion loss of the taper structure.
In \Fref{fig:S4} we show transmission through the sample measured in the main text, de-embedded by referencing a cryogenic through calibration performed in the prior cooldown.
We find that the broad spectral features match reasonably well with simulated insertion loss of the taper structure.
In practice we find the parasitic modes around 100~GHz from the indium mounting regions vary from sample to sample, resulting in a wider frequency distribution and higher loss, leading to broader and shallower features.
With the background spectral features determined, the remaining sharp dips in transmission distinguish the readout resonator frequencies (between 90--98~GHz).

\subsection{Two-tone Characterization}
\begin{figure*}[htb]
\centering
\includegraphics[width=6.0in]{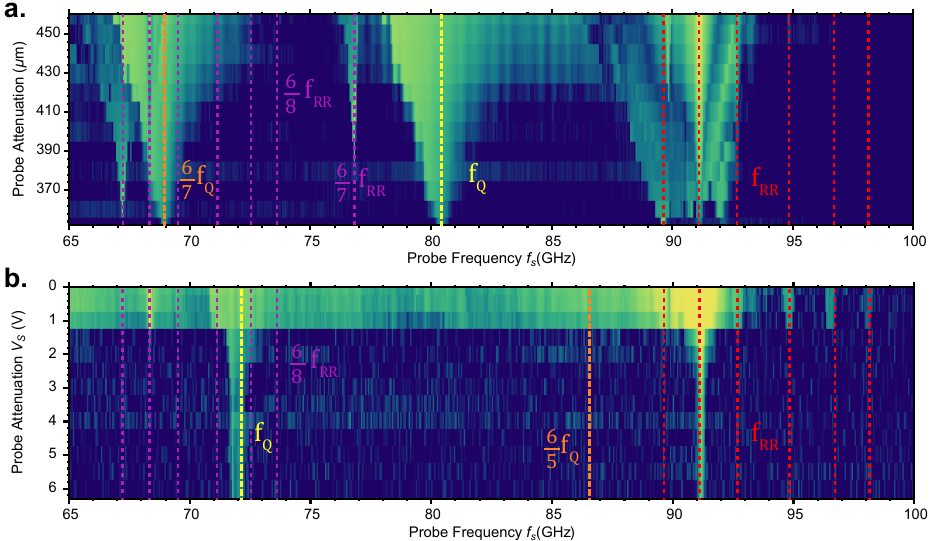}
\caption{
Broadband two-tone spectroscopy.
(a-b) Deflection of the readout resonator as a function of applied probe frequency, shown for varying attenuation conditions. For both cases the vertcial axis corresponds to increasing probe power. 
Due non-negligible cross-coupling and the harmonics involved in measurement, multiple spectral features appear.
These can be assigned to the readout resonators (red), qubit (yellow) and their harmonics (purple and orange).
The dataset in (a) is measured approximately one year prior to the experiments in (b) and the main text, in between which the sample was re-etched before remeasuring.
This predominantly affects the junction, shifting qubit frequencies down by about 8~GHz.
\label{fig:S5}}
\end{figure*}
An important step in characterizing the qubit-resonator system is spectroscopically determining qubit transition frequencies. 
While these measurements were performed on each qubit-readout resonator pair, we focus on qubit 2, which showed the narrowest transition linewidths while still falling within the 70-85~GHz range of the pulsed measurement hardware.
Similar to the measurements in Fig. 2b, we first search for the qubit in the full operating band: monitoring transmission at the readout resonance $f_R$ while simultaneously applying a second probe signal $f_S$ to reveal the excitation spectrum of the coupled system shown in \Fref{fig:S5}.

Along with signals at the qubit frequency (detailed in Fig. 2b) we also observe several other features in the spectrum.
Due to inherited resonator nonlinearity from the coupled qubit \cite{reed2010snap}, and stray cross-coupling between neighboring resonators, signals are also observed at the readout resonator frequency and its neighbors.
This spectrum is also complicated by our use of harmonic multipliers to generate both the local oscillator for the readout mixer $f_{RLO}$ and probe signal $f_S$, producing integer $n$ harmonics of the generating frequency:
\begin{equation}
    f_{S,n} = \frac{n}{6}f_{S},\quad f_{RLO,n} = \frac{n}{6}f_{RLO}
\end{equation}
While lower frequencies are suppressed by the natural evanescent cutoff of the rectangular waveguide, the higher frequency $n>4$ harmonics are still prominent.
Because of this, the down-mixed readout measurement is sensitive to multiple millimeter-wave frequencies, all of which can down-mix to the same idler frequency (IF) as illustrated in \Fref{fig:S3}a.
This results in spurious signals whenever a probe harmonic lines up with a measurement frequency, however have very narrow linewidth so can be eliminated from the experiment by averaging the measurement or using coarse or non-integer frequency sweep steps.

Having eliminated direct conversion spurs, the features in \Fref{fig:S5} must occur when either the qubit or readout resonator are excited by the probe frequency or one of its harmonics:
\begin{equation}
    f_S =\frac{6}{n}f_{RR_i}, \quad
        \frac{6}{n}f_{Q_j}
\end{equation}
Allowing for cross-coupling, when measuring a given readout resonator we could also expect signals from its neighboring readout resonators $RR_i$, and potentially even neighboring qubits $Q_j$ (though in practice cross-coupling $i\neq j$ between mismatched qubits and resonators is significantly weaker).
With these rules established, the features in the spectra shown in \Fref{fig:S5} can be assigned to respective excitations, identifying the true qubit transition frequency.

\section{Power Calibration using Qubit Spectroscopy}
\begin{figure*}[htb]
\centering
\includegraphics[width=5.0in]{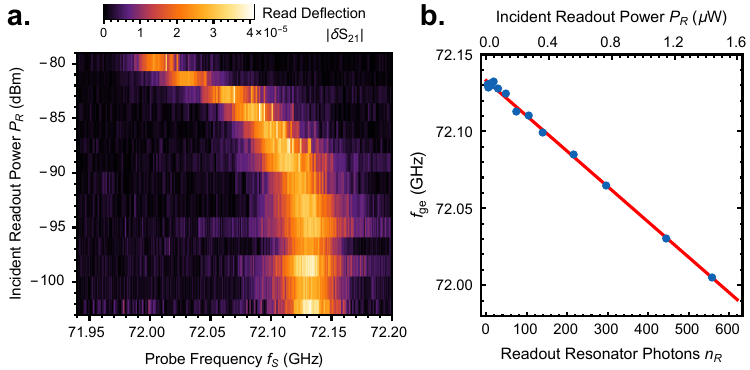}
\caption{
AC Stark shift calibration.
(a)Deflection of transmission through the readout resonator at its resonant frequency measured as a function of applied probe tone frequency $f_S$.
When $f_S$ is resonant with the qubit ground state transition a peak is observed.
As the readout power is increased, the qubit transition frequency shifts down due to the cross-Kerr interaction $\chi$ between the qubit and readout resonator.
(b) The qubit frequency as determined by the peak location from (a) is linear in the number of intracavity readout resonator photons $n_R$, providing a calibration for the incident readout resonator power.
\label{fig:S6}}
\end{figure*}
\begin{figure*}
\centering
\includegraphics[width=6.5in]{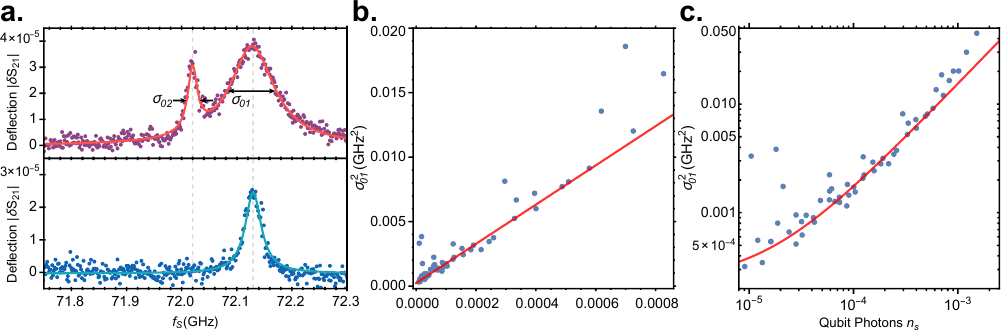}
\caption{
Linewidth power broadening. (a) The feature corresponding to the ground--excited state transition in the readout resonator deflection measurements has a gaussian profile, allowing us to fit the linewidth of the transition.
Measurements on the top and bottom are shown for drive strengths corresponding to $2.12\times10^{-4}$ and $3.02\times10^{-5}$ qubit photons respectively.
For sufficiently high powers (top), the edges of the transition overlap with the next two-photon transition, so the model includes both of these transitions.
(b) The fitted $0$--$1$ transition linewidths $\sigma_{01}$ are plotted as a function of drive photon number.
The square of the linewidth is determined by the qubit dephasing time, and increases linearly with drive power, with a rate set by the qubit decoherence time.
\label{fig:S7}}
\end{figure*}

Having identified its transition frequencies, the qubit provides an excellent opportunity for calibrating millimeter-wave power.
The qubit-resonator dispersive shift $\chi$ can be determined from the qubit frequency, its anharmonicity, the bare coupling strength and resonator detuning.
The dispersive shift can also be measured directly by monitoring the frequency shift of the qubit transition with respect to the average number of photons populating the readout resonator:
\begin{equation}
    f_{ge}(n_R) =f_{ge,0} - \chi n_R
\end{equation}
To investigate this, we repeat the two-tone experiment in Fig. 2b as a function of probe frequency at a very low probe power.
The results are summarized in \Fref{fig:S6}, in which we observe that as the readout power increases the qubit transition $f_{ge}$ decreases.
Transforming the readout power into resonator photon number we verify that the frequency shift is linear, and based on the dispersive shift calculated from previously measured values, determine the effective signal attenuation.
We find that results of this method agrees with the values calculated using transmission measurements calibrated with a power meter within 5~dB.
This calibration provides a translation between applied power, incident power and readout resonator photon number based on the qubit response, as shown in \Fref{fig:S6}b.

\section{Qubit Coherence from Power Broadening}

The linewidth of the fundamental qubit transition is expected to scale with applied power \cite{schuster2005ac} which is proportional to $n_s g^2$, where $g$ is the bare coupling strength and $n_s$ denotes the average number of qubit photons at a particular drive strength.
The half-width half-maximum $\sigma$ of the transition feature takes the following expression \cite{schuster2005ac}:
\begin{equation}
    2\pi \sigma = \frac{1}{T_2'} = \sqrt{\frac{1}{T_2^2}+n_S g^2\frac{T_1}{T_2}}
\end{equation}
Which depends on the qubit dephasing and relaxation rates.
In \Fref{fig:S7}b-c we plot $\sigma^2$ as a function of the calibrated qubit photon number $n_s$, finding that the square of the linewidth has the expected linear relationship.
The intercept and slope allow us to estimate a dephasing rate of $T_2=20.9~$ns and relaxation time of $T_1=47.3~$ns respectively.
We note that in this method, the dephasing time is only calculated from fitting the linewidth, however the relaxation time requires a sequence of additional measurements, making the relaxation time estimate significantly less accurate.

\section{Frequency Dependence of Rabi Oscillations}
\label{appendixF}
\begin{figure*}[htb]
\centering
\includegraphics[width=6.5in]{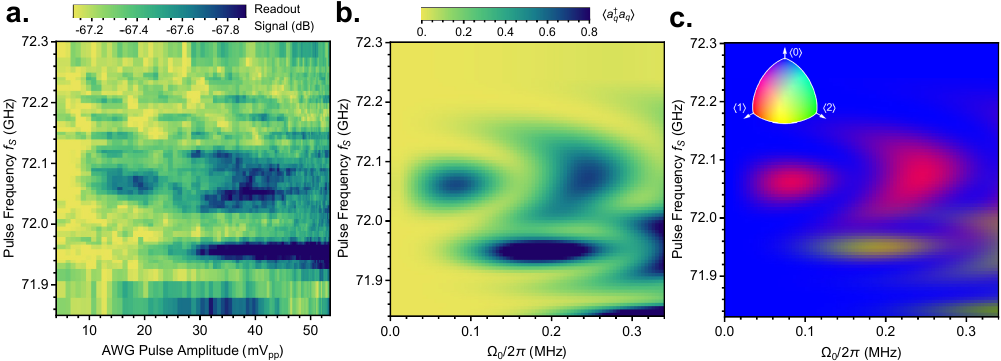}
\caption{
Frequency-dependence of Rabi oscillations. (a) Qubit state oscillations as a function of pulse amplitude ($\Omega_0$) and frequency, with master equation simulations of the corresponding experiment shown in (b-c). 
Here the pulse length is 4~ns, with $\sigma=2~$ns (with higher attenuation compared to Fig. 3).
A chevron pattern of fringes is visible near the $f_{01}$ transition frequency, which has been Stark-shifted by leaking local oscillator signal.
Since the Rabi rates are comparable to the qubit anharmonicity, oscillations between the higher energy levels also appear at higher-powers.
These features are prominent in the readout resonator signal, which is more sensitive to higher qubit energy states.
(c) Simulated populations of the first three energy levels of the qubit help separate and identify the fringes corresponding to each qubit transition.
\label{fig:S8}}
\end{figure*}
In Fig 3, we study Rabi oscillations in time domain, finding that the generalized Rabi frequency decreases while the oscillation amplitude grows as the pulse frequency approaches the qubit transition frequency $f_{01}$.
Notably this optimum frequency is lower than from the transitions measured with continuous-power two-tone spectroscopy in Fig. 2, likely due to the ac Stark shift from the significant millimeter-wave leakage incident on the qubit (see \Fref{fig:S3}b).
To study the frequency-dependence of the Rabi oscillations, we repeat the experiment in Fig. 3 as a function of pulse amplitude and frequency
(this experiment can be run far more rapidly by holding the qubit IF frequency fixed, which alleviates the need for re-uploading pulse waveforms into the AWG).
The results are shown in \Fref{fig:S8}a, which captures the first two fringes of the expected chevron pattern but centered at a lower frequency than the $f_{01}$ measured with two-tone spectroscopy, and consistent with the discrete-frequency Rabi oscillation experiments in Fig. 2.
We note that while amplitude nonlinearity (detailed in \Fref{fig:S3}c) was corrected for in this experiment, the actual pulse power depends significantly on frequency (as illustrated in \Fref{fig:S3}b), which amounts to varying horizontal skew in the experimental data.

In addition to the chevron pattern, we also observe two features at lower frequencies corresponding to oscillations between the higher energy levels.
To better understand these features, we perform master equation simulations of the ac-Stark-shifted qubit Hamiltonian \cite{Schneider2018multiStark} excited by the same pulse conditions.
Since the dispersive shift is small relative to the readout resonator linewidth, we approximate the measured readout shift to be proportional to the expected value of the qubit state
$|\delta S_{21}|\propto \int\chi\langle a^\dag_q a_q\rangle dt$ integrated over the readout duration (20~ns).
The results are summarized as a function of maximum drive strength $\Omega_0$ in \Fref{fig:S8}b, which shows fairly good agreement with the experimental data, capturing the pronounced features corresponding from oscillations between the higher energy levels.
These features are also lower than the frequencies measured in two-tone experiment \cite{Schneider2018multiStark}.
For more clarity, we can further separate the simulated qubit occupation in \Fref{fig:S8}b into individual qubit state probabilities, as shown in \Fref{fig:S8}c.
This confirms that the dominant Rabi oscillations on-resonance with the shifted qubit frequency involve only the $|0\rangle$ and $|1\rangle$ states, while the lower-frequency fringes excite population in the higher-energy $|2\rangle$ state.

\section{Qubit Loss Sources}
\subsection{Leakage into Tapered Coupler Modes}
\begin{figure*}
\centering
\includegraphics[width=5.0in]{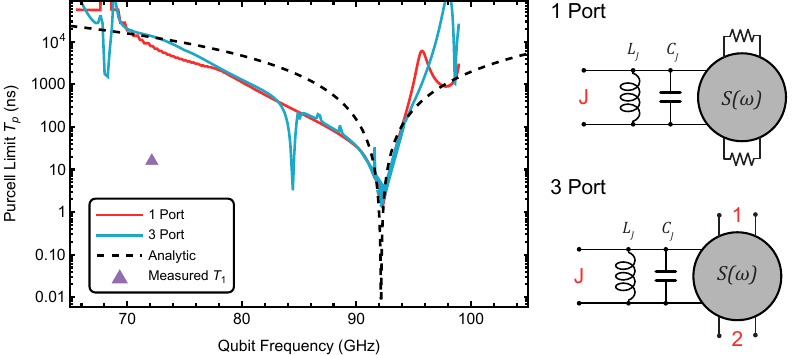}
\caption{
Purcell limit in the taper structure.
Calculated Purcell lifetimes are shown for analytic expression (dashed) which assumes the simplified circuit model (Fig. 1e) as well as purcell limits calculated using the black box method \cite{nigg2012blackbox} using the admittance from finite element method simulations of the entire taper structure.  To approximate energy radiation from the input and output waveguide ports, we either replace the input and output waveguides with guided wave boundaries (1-port) or simulate the full 3-port parameters. We find the latter provides improved mesh distribution highlighting finer mode structure details. In both cases the junction impedance $L_J$, $C_J$ is varied to calculate the effective qubit mode frequency.
All three methods estimate significantly higher limits on qubit coherence than our measurements.
\label{fig:S9}}
\end{figure*}
Near the qubit frequency, the tapered coupling structure \cite{anferov2024taper} does not have perfect transmission, reflecting some signal back.
This allows the coupling structure to act as a mild Purcell band-pass filter \cite{purcell1995spontaneous,Yan2023purcellReadout}, expected to reduce qubit decoherence from spontaneous emission into the input or output waveguide.
Since some of the electromagnetic eigenmodes of the taper structure span the entire chip however, the qubit can also spontaneously decay through its weak cross-coupling with the taper modes, which dissipate directly into waveguide fields.
The wavelength-scale geometry of the taper structure makes a circuit model of the filter \cite{Yan2023purcellReadout} less accurate, so we instead rely on finite-element simulations of the entire chip and waveguide assembly to establish network parameters.
To estimate leakage into the waveguide, we follow the black box method \cite{nigg2012blackbox}. 
The simulated network provides the input admittance $Y(\omega)$ as seen by the qubit Josephson junction, which when recombined with the junction admittance $Y_J = - i\omega C_J + i/\omega L_J$ determines the mode frequencies from the admittance zeroes.
The Purcell limit can also be calculated from the admittance and its frequency derivative at the qubit mode frequency $\omega_q$ \cite{nigg2012blackbox}:
\begin{equation}
    T_P = \frac{1}{2}\frac{\text{Im}Y'(\omega_q)}{\text{Re}Y(\omega_q)}
\end{equation}
We repeat this calculation for different junction areas (rescaling $L_j$ and $C_J$ appropriately), and summarize the results in \Fref{fig:S9} along with the analytic Purcell limit \cite{purcell1995spontaneous,Houck2008purcell,Sete2014purcellnonlin} assuming the readout resonator directly couples to the waveguide.
We find that replacing guided wave boundaries on the input and output waveguides with a three-port simulation provides a more detailed estimate since this better captures the full mode structure of the taper (as the simulation mesh refinement follows energy flow, which is otherwise only centered on the qubit).
Notably the relaxation times simulated with this method can be nearly an order of magnitude lower than the analytical estimate, with the three-port results highlighting a particuarly leaky mode near 84~GHz.
We note that all of these methods yield significantly higher relaxation rate limits than measured experimentally, suggesting that spontaneous emission is not the dominant source of decoherence. Nevertheless these simulations provide more insight into to the mode structure of the tapered coupler assembly, and may be useful for better understanding other potential sources of decoherence such as measurement-induced loss 
\cite{dumas2024ionization,Shillito2022ionization}.

\subsection{Decoherence from Junction Materials}
\begin{figure*}
    \centering
    \includegraphics[width=6.6in]{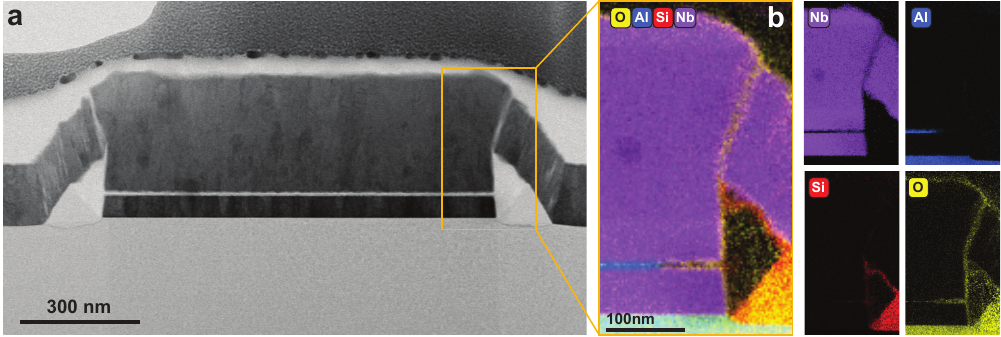}
    \caption{
    Cross-sectional materials analysis. (a) Bright field STEM image overview of the trilayer junction. Yellow rectangle shows structure where the EDS was taken. (b) Composite Energy Dispersive Spectroscopy (EDS) image showing normalized weight percentage for O, Al, Si, and Nb, with individual element density maps shown in their respective color on the right. }
    \label{fig:S10}
\end{figure*}

The electric field concentration in and around the trilayer junction is orders of magnitude higher than anywhere else in the qubit structure, making the qubit especially sensitive to the presence of lossy materials near the junction barrier \cite{anferov2023nbjj}.
In particular niobium oxides \cite{verjauw2021nbOxide,premkumar2021nbOxide} and un-removed spacer material \cite{chang2014striploss,Shan2024tlsloss} are both likely sources of decoherence channels from two level systems.
To further investigate materials present near the junction barrier, we perform transmission electron microscopy (TEM) analysis on a cross-sectional lamella cut from the junction, summarized in \Fref{fig:S10}. 

The lamella for TEM analysis is prepared by standard focused ion beam (FIB) in-situ lift-out technique \cite{schaffer2012sample} in a FEI Helios NanoLab 600i DualBeam (FIB) system. 
The sample is first coated with carbon and then  with gold to enhance conductivity and protect the top surface before loading into the FIB. The sample is then coated with platinum using the gas injection system to offer more protection under FIB milling. 
The lamella is extracted in-situ using the OmniProbe AutoProbe 200, and later welded onto a standard copper TEM grid for analysis. 
TEM and scanning transmission electron microscopy (STEM) analysis are performed in a 300~kV Thermo Fisher Spectra 300 microscope capable of aberration correction for both TEM and STEM. Data is collected and analyzed using Velox software from Thermo Fisher. 

The resulting cross-sectional STEM image is shown in \Fref{fig:S10}a, with the lamella oriented in the lateral direction from perspective of Fig. 1. 
Importantly we observe that a significant portion material is present in the triangular spacer regions, as well as potential fractures in the wiring layer.
To better differentiate between different materials, we also probe the elemental composition using energy dispersive x-ray spectrocopy (EDS) to generate an elemental map of the junction structure as shown in \Fref{fig:S10}b.
As expected, we observe high Al and O concentration in the sapphire substrate, with high Nb concentration in the junction electrodes. 
However, the elemental distribution also confirms that the amorphous SiO$_2$ spacer is only about halfway removed, and reveals the presence of Si-rich residue on the underside of the wiring layer: both of which significantly limit the junction loss performance \cite{anferov2023nbjj}. 

Furthermore, we find several niobium oxide regions in the junction, which may contribute to qubit decoherence.
Near the edges of the junction, the Al signal decreases, consistent with the previously observed $\sim80~$nm dimension reduction \cite{anferov2023nbjj} resulting from the Al junction layers being partially dissolved during the spacer removal etch. 
In these locations where the Al is removed, the exposed Nb layers form surface oxides, visible in \Fref{fig:S10}b: this niobium oxide is located near the junction interface where its electric field is strongest, and likely also adversely affects the junction's coherence properties.
This oxide growth could be minimized by further reducing the aluminum etch rate (such as with vapor-based spacer removal methods).
Finally we observe internal fractures close to the top of the wiring layer, which have subsequently oxidized. We hypothesize that surface tension from the solvent cleaning steps following the spacer removal etch may pull the Nb bridge downwards, forming stress fractures in the wiring layer that subsequently oxidize.
While this damage could have also occurred after fabrication, the presence of the fractures suggests the possibility of dissipation from imperfect superconducting contacts in the qubit junction.
These effects could be alleviated by reducing surface tension in the final drying process either with a critical point solvent, or by vapor-based spacer removal.

Our TEM cross-sectional study of the trilayer junction reveals several structural features that could be dominant limits of decoherence of qubits. First, we see the residual SiO$_2$ in the spacer region. Second, there is niobium oxide growing near the junction interface where Al is removed. Finally, the niobium oxide gap's formation on the Nb top layer can further affect the coherence of junction. 

\bibliography{thebibliography}
\end{document}